\documentclass[aps,twocolumn,floatfix,amsmath,amssymb]{revtex4}
\usepackage{graphicx}
\usepackage{dcolumn}
\usepackage{bm}

\begin{document}

\title{Rydberg trimers and excited dimers bound by internal quantum reflection}

\author{V. Bendkowsky}
	\email{v.bendkowsky@physik.uni-stuttgart.de}
\author{B. Butscher}
\author{J. Nipper}
\author{J. Balewski}
\author{J. P. Shaffer}
  \altaffiliation{University of Oklahoma, Homer L. Dodge Department of Physics and Astronomy, 440 W. Brooks St., Norman, Oklahoma 73019, USA}
\author{R. L\"{o}w}
\author{T. Pfau}
\affiliation{
5. Physikalisches Institut, Universit\"{a}t Stuttgart, Pfaffenwaldring 57, 70569 Stuttgart, Germany.}

\author{W. Li}
\author{J. Stanojevic}
\author{T. Pohl}
	\email{tpohl@pks.mpg.de}
\author{J. M. Rost}
\affiliation{
Max-Planck-Institut f\"{u}r Physik komplexer Systeme,
Noethnitzer Str. 38, 01187 Dresden, Germany.}

\date{\today}

\begin{abstract}
Quantum reflection is a pure wave phenomena that predicts reflection of a particle at a changing potential for cases where complete transmission occurs classically. For a chemical bond, we find that this effect can lead to non-classical vibrational turning points and bound states at extremely large interatomic distances. Only recently has the existence of such ultralong-range Rydberg molecules been demonstrated experimentally.  Here, we identify a broad range of molecular lines, most of which are shown to originate from two different novel sources: a single-photon associated triatomic molecule formed by a Rydberg atom and two ground state atoms and a series of excited dimer states that are bound by a so far unexplored mechanism based on internal quantum reflection at a steep potential drop. The properties of the Rydberg molecules identified in this work qualify them as prototypes for a new type of chemistry at ultracold temperatures.
\end{abstract}

\pacs{Valid PACS appear here}
\maketitle

The development of techniques to manipulate ultracold atoms has opened up new avenues for molecular physics. The recent experimental success in producing cold ground state molecules \cite{nom08,dhg08,dgr08} is paving the way towards studying chemistry at ultralow  temperatures \cite{kre06}. This regime also offers unprecedented opportunities for the control of large-scale molecules involving highly excited states \cite{ost09,bbn09}, which hold great promise for such applications due to the high sensitivity of Rydberg atoms to external fields \cite{lps06}.

A particular class of Rydberg molecules arises from binding between a Rydberg and a ground state atom, which is based soley on low-energy collisions between the Rydberg electron and the ground state atom \cite{gds00}. Compared to the large de Broglie wavelength of the electron, the ground state atom represents a point-like perturbation which probes directly the Rydberg atom's wavefunction. Consequently, the molecular potential shows the oscillatory character of the corresponding probability density. Owing to the large size of Rydberg atoms, binding occurs at extremely long range on the order of several thousand Bohr radii, $a_0$.

Recently, we have provided experimental proof for the existence of these ultralong-range molecules in an ultracold rubidium gas \cite{bbn09}. A simple model with an adjustable s-wave scattering length was found to reproduce the binding energies of the vibrational ground states for a range of principal quantum numbers.

In this work, we demonstrate the existence of several additional molecular lines. The observed spectra can be adequately described with Fermi's pseudopotentials for local Rydberg electron scattering off the perturbing ground-state atom \cite{fermi,omont}, but within a non-perturbative Green's function approach which accounts for the strong collisional couplings between Rydberg state manifolds.
Despite the lack of an inner potential barrier, the calculated excited states have the peculiar property to avoid smaller interatomic distances. As we will show, this is due to internal quantum reflection at a shape resonance of electron-atom scattering. 
In addition to reproducing all previously observed lines \cite{bbn09}, the calculations predict additional dimer and trimer states that are observed in our new experiments.

\section{The molecular potential}

\begin{figure}[t!]
\begin{center}
\resizebox{0.95\columnwidth}{!}{\includegraphics{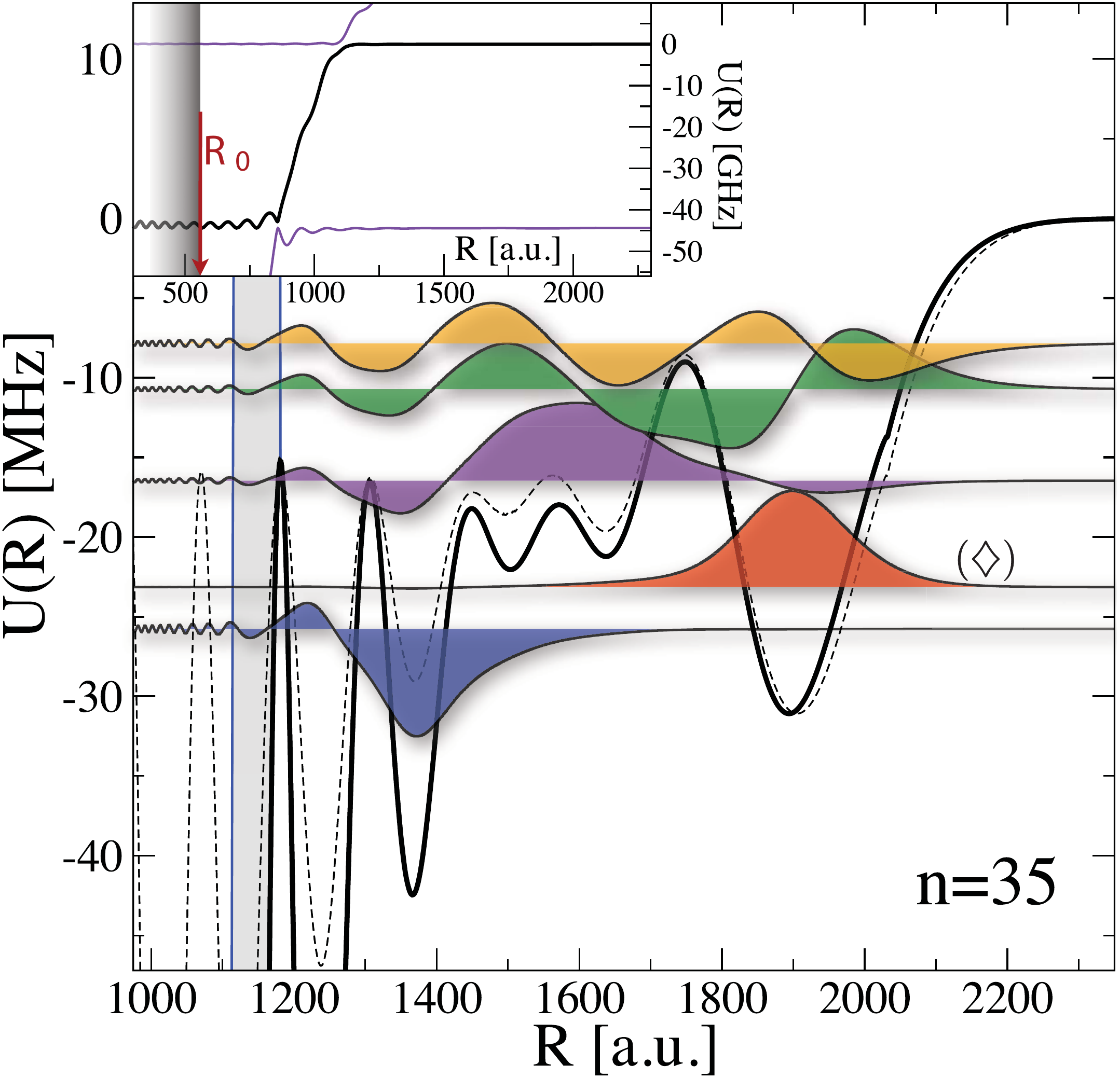}}
\caption{\label{fig2}Molecular potential curve for $n=35$. Results obtained from the Green's function calculation (solid line) and from the model potential Eq.(\ref{fermi}) (dotted line) with an effective scattering length $A_s^{\rm (eff)}=-19.48\,a_0$. The grey area around $R\approx 1200\,a_0$ marks the innermost "badland" region defined by $d\lambda/dR>1$ \cite{cht97}. The baseline of the molecular wave functions corresponds to their respective vibrational energies. The inset shows the rapid potential drop at the avoided crossing and illustrates the variable inner boundary condition at a distance $R_0$.}
\end{center}
\end{figure}

The state of the molecular Rydberg electron is determined by both long-range attraction to its parent ion as well as the short-range interaction   with the nearby ground state atom. This vast length scale disparity poses considerable problems for current quantum chemistry approaches, which can be elegantly resolved through Fermi's pseudopotential approximation \cite{gds00}.
We first apply the Born-Oppenheimer approximation and a single-active particle treatment of the excited electron to obtain the following  Hamiltonian (in atomic units)
\begin{equation}\label{ham}
H=-\frac{\nabla^2}{2}+V_{\rm eA^+}(|{\bf r}-{\bf R}|)+V_{\rm eA}({\bf r})\equiv H_{{\rm A}}+V_{\rm eA}({\bf r})
\end{equation}
of the molecular Rydberg electron. Here, $\hat{H}_{{\rm A}}$ is the Hamiltonian of the unperturbed Rydberg atom, whose center is located at ${\bf R}$, and where the molecular axis ${\bf R}$ also defines the quantization axis. $\hat{V}_{\rm eA}({\bf r})$ accounts for the interaction of the Rydberg electron with the perturbing ground state atom, which is described by a Fermi-type pseudopotential for electron atom scattering \cite{fermi,omont}, including the first two $l=0,1$ partial waves.

Within the Fermi model \cite{fermi}, the molecular interaction of low-$\ell$ Rydberg states is commonly described by the simple first-order perturbative potential \cite{fermi,omont} 
\begin{equation} \label{fermi}
U^{(1)}(R)=2\pi A_{\rm s}|\psi_n(R)|^2+6\pi A_{\rm p}^3|\nabla \psi_n(R)|^2
\end{equation}
where $\psi_n$ is the unperturbed electronic Rydberg wave function and $A_{\rm s(p)}(k)$ is the energy dependent s(p)-wave scattering length. The electron momentum $k$ depends on the Rydberg energy $E_n$ through the quasiclassical relation $E_n=k^2/2-1/R$. This approach yields an intriguingly simple and intuitive description of the molecular interaction and is typically employed to describe molecules formed with energetically well isolated $ns$ Rydberg states \cite{gds00,kcf02}.
Even for this case, however, Eq.(\ref{fermi}) turns out to be inaccurate at the level of accuracy of the experimental data. Hence, we  performed non-perturbative Green's function calculations \cite{akm84} of the molecular potentials (see Methods). 

As an example, we compare the calculated potential energy curves $U(R)$ with the first-order perturbative potential $U^{(1)}(R)$ for $n=35$ in Fig.~\ref{fig2}. 
As is known for high-$\ell$ Rydberg molecules, Eq.(\ref{fermi}) fails in the vicinity of the p-wave shape resonance \cite{hgs02,ghc06} because the resonance leads to an avoided potential crossing. This curve crossing is crucial for determining the vibrational spectrum of the molecule found in our experiments. Unexpectedly, $U^{(1)}$ also falls short of more accurate calculations around the outer potential well at $R=1900\,a_0$. 

One can, nevertheless, obtain an approximate description of the strong molecular resonance, localized in the outermost minimum of the potential energy curve, by using an effective, properly adjusted s-wave scattering length $A_s^{\rm (eff)}(k=0)$ with Eq.(\ref{fermi}) (see dotted line in Fig.~\ref{fig2}). The corresponding vibrational energy for all principal quantum numbers $n=35,36,37$ is well reproduced for an effective scattering length of $A_s^{\rm (eff)}=-19.48\,a_0$. While this value is $20\%$ larger than the actual value of $A_{\rm s}(k=0)$ (see Sec.~II), it may serve as an effective parameter or characteristic of a given atomic species that yields the respective $n$-dependence of the strongest molecular resonance, based on the simple model potential Eq.(\ref{fermi}).

\section{Experimental results}

\begin{figure}[t!]
\begin{center}
\includegraphics{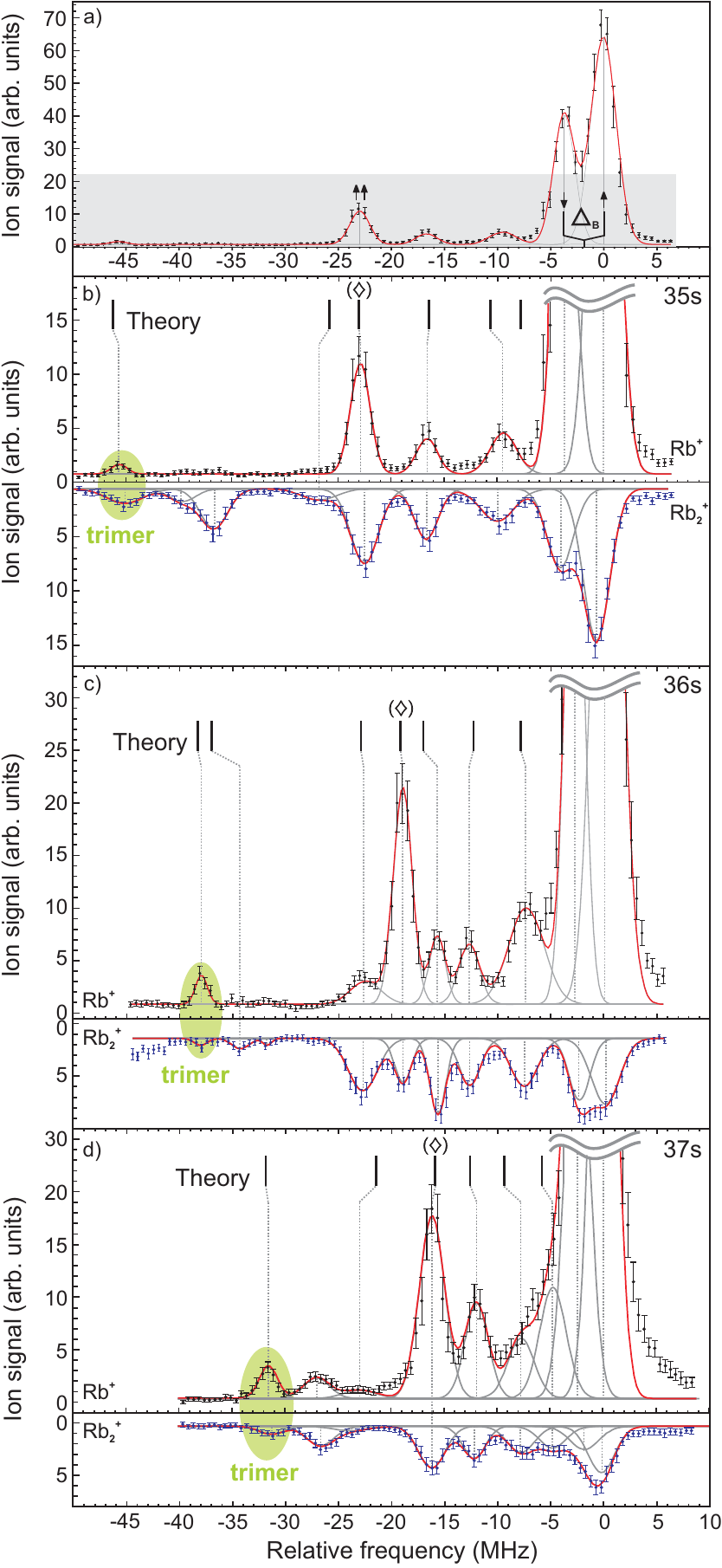}
\caption{\label{fig1}Spectra for principal quantum numbers $n=35-37$. a) Full spectrum for Rb(35s) and illustration of the atomic Zeeman shift $\Delta_B$. b)~-~d) Magnified spectra of the states $35s$ to $37s$. Frequencies in all spectra are measured with respect to the atomic transition $5s_{1/2}\rightarrow ns_{1/2}$. The upper parts show the atomic ion spectrum $\mbox{Rb}^+$(black) and the lower ones the spectra of the molecular ions $\mbox{Rb}_2^+$ (blue). Calculated binding energies (Tab.~\ref{tab1}) are indicated by vertical lines.}
\end{center}
\end{figure}

To create the ultralong-range Rydberg molecules, we prepare a cold sample of $^{87}\mbox{Rb}$ atoms in the ground state $5s_{1/2}, F=2, m_F=2$. The $ns_{1/2}$ Rydberg state of principal quantum numbers, $n=35-37$, is addressed by two narrow-band lasers (see Methods). The vibrational bound dimer and trimer states can be created directly via photoassociation if the detuning of the excitation light from the atomic Rydberg state matches the molecular binding energy, $E_B$. Therefore, the molecular states appear in the spectra on the red side of the atomic Rydberg line, as shown in Fig.~\ref{fig1}. The uppermost graph (a) shows the full $\mbox{Rb}(35s)$ spectrum which consists of two atomic states $m_s=+1/2~(\uparrow)$ and $m_s=-1/2~(\downarrow)$, separated by the Zeeman splitting $\Delta_B$ in the magnetic trap, and several molecular lines. Due to the spin polarization in the magnetic trap $(m_F=+2)$, only the molecular triplet states $(\uparrow \uparrow)$ are excited. The atomic state $ns_{1/2},m_s=+1/2$ and the molecular state $^3\Sigma(ns-5s)$ have the same Zeeman shift (see inset) and thus the binding energy $E_B$ can be directly measured as the energy difference between the corresponding lines in the spectrum. Note that in our previous work \cite{bbn09} the binding energies $E_B$ were defined differently. Fig.~\ref{fig1} (b)-(d) show the $35s$ to $37s$ spectra at high resolution.  

After field ionization, we observe both an atomic ion signal corresponding to $\mbox{Rb}^+$ and a molecular ion signal corresponding to $\mbox{Rb}_2^+$ which can be clearly identified in the time-of-flight spectra. The $\mbox{Rb}^+$ and $\mbox{Rb}_2^+$ spectra predominantly show the same features but with different intensity distributions. A few deep bound states exclusively appear in the Rb$_2^+$ spectrum, but can be assigned as dimer states by our theoretical model. The fact that the strengths of the two ion signals are comparable for the molecular states is a clear indication that $\mbox{Rb}_2^+$ is the product of an additional decay that  occurs for the molecular states. To obtain maximum spectral information, we use both the Rb$^+$ as well as Rb$_2^+$ for the following analysis of the molecular data.

To assign the observed lines based on the described potential calculations, we employ the ab-initio data of \cite{kcf02,hgs02} for the s-wave scattering phaseshift $\delta_{0}(k)$. Since the available p-wave scattering data \cite{bt00,kcf02} does not yield an accurate description of the measured spectra, we use the modified effective range expression \cite{msr61} for the corresponding phaseshift $\delta_1$. This three-parameter fitting procedure allows the assignment of most of the observed lines and reproduces the resonance frequencies with a remarkable accuracy. The results of the calculation are included in Fig.~\ref{fig1} and compared to the measurements in Tab.~\ref{tab1}.
The strongest molecular line, labeled by $(\diamondsuit)$, in each spectrum corresponds to the dimer state localized in the outermost potential well (Fig.~\ref{fig1}) which was assigned as vibrational ground state $\nu=0$ in \cite{bbn09}. The present study allows to identify several additional lines as excited dimer states. In particular, the theory predicts a state which is even deeper bound than the state in the outermost potential well, $(\diamondsuit)$. This state is localized around $R\approx1400\,a_0$ for 35s (Fig.~\ref{fig1}) and is assigned in the measured spectra.
In addition to the dimer lines, we observe two resonances at deeper binding energy in the experiment. The lowest one systematically appears at twice the binding energy of the outermost localized dimer state $(\diamondsuit)$. Our Green's function calculations show that for such large internuclear separations, the binding energy for two simultaneously bound ground state atoms is approximately given by the sum of the two binding energies. This approximation is a result of the weak interaction between the ground state atoms and the large geometric phase space of the trimer molecule. The observed spectral feature provides clear evidence for the direct photoassociation of a long-range triatomic molecule \cite{liu06,liu09}, in which two ground state atoms are simultaneously bound to the Rydberg atom via their interaction with the Rydberg electron. The remaining unassigned lines may indicate that excited states of these exotic trimers are also formed in our experiments.

The agreement between theory and experiment confirms existing s-wave scattering data, predicting a zero energy s-wave scattering length of  $A_s(k=0)=-16.05\,a_0$ \cite{bt00,kcf02}. In addition, the described fitting procedure allows to extract p-wave scattering parameters, and  yields a zero energy scattering length of $A_p(k=0)=-21.15\,a_0$, which is well within the range of previous calculations. Our extracted value for the shape resonance energy $E_{\rm res}=23.6$ meV constitutes the first experimental determination of this quantity and agrees well with theoretical predictions \cite{jb82,bt00}.

\begin{table*}[t]
\begin{center}
\begin{tabular}{|c|c|c|c||c|c|c|c||c|c|c|c|}
\hline 
35s &\multicolumn{2}{c|}{measurement} & theory & 36s &\multicolumn{2}{c|}{measurement} & theory & 37s &\multicolumn{2}{c|}{measurement} & theory \tabularnewline
state &Rb$_2^+$ & Rb$^+$ & ~ & state &Rb$_2^+$ & Rb$^+$ & ~& state &Rb$_2^+$ & Rb$^+$ & ~ \tabularnewline
\hline 
\hline 
~ & $/$ & $/$ & -7.8 & ~ & $-7.7(6)$ & $-7.4(6)$ & $-7.8$& ~ & $-4.8(10)$ & $/$ & $-5.8$ \tabularnewline
\hline 
~& $-9.9(7)$ & $-8.9(7)$ & $-10.7$ & ~ &$-12.8(9)$ & $-12.8(9)$ & $-12.3$ & ~&$-7.8(10)$ & $/$ & $-9.4$ \tabularnewline
\hline
&$-16.6(5)$ & $-16.5(5)$ & $-16.5$  & ~ &$-15.8(8)$ & $-15.8(5)$ & $-17.0$ &~&$-12.2(5)$ & $-12.0(5)$ & $-12.6$ \tabularnewline
\hline
$ \diamondsuit$ & $-22.4(5)$ & $-22.8(5)$ & $-23.1$ & $ \diamondsuit$ &$-19.0(5)$ & $-19.0(5)$ & $-19.2$& $ \diamondsuit$ &$-16.2(5)$ & $-16.2(5)$ & $-15.9$ \tabularnewline
\hline
~ & $-26.3(7)$ & / & $-25.9$ & ~ & $-22.8(5)$ & $-22.6(8)$ & $-22.9$ & ~ &$-23.8(15)$ & $-23.0(15)$ & $-21.5$ \tabularnewline
\hline
&$-36.6(5)$ & / & /& ~ & $-31.9(10)$ & / & / & ~ & $-26.7(10)$ & $-27.1(10)$ & / \tabularnewline
\hline
&$-39.6(12)$ & / & / & ~ & $-34.4(10)$ & / & $-37.0$ & ~ & / & / & / \tabularnewline
\hline
trimer&$-45.0(8)$ & $-45.5(8)$ & -46.3& trimer&$-38.1(8)$ & $-37.9(5)$ & -38.3& trimer& $-31.3(8)$ & $-31.6(8)$ & $-31.9$ \tabularnewline
\hline
\end{tabular}
\end{center}
\caption{\label{tab1}Observed and calculated molecular resonances. The frequencies are given in MHz. Errors $(2\sigma)$ of the measured binding energies are determined by the fit of the spectra.}
\end{table*}

\section{Binding mechanisms}

For a pure s-wave potential the outermost localized state $ (\diamondsuit)$, for $n=35$ at $-23~$MHz, constitutes the true vibrational ground state \cite{bbn09} of the molecule. 
A more accurate determination of the molecular potential requires the inclusion of p-wave scattering and in particular the shape resonance in the p-wave scattering cross-section of rubidium \cite{bt00,jb82} has to be taken into account, which leads to an avoided crossing between adjacent molecular potential curves around $\sim 1000\,a_0$ (see inset of Fig.~\ref{fig1}).
This has a dramatic effect on the vibrational structure and results in additional molecular states that are delocalized over several $100~a_0$ with vibrational energy that lies above the potential maxima inside the attractive potential well. Although energetically possible, these states have only negligible probability density at interatomic distances smaller than $\sim$ 1000$a_{0}$. As there exists no inner potential barrier that would facilitate a common molecular binding of these excited vibrational states, the binding has to arise from a fundamentally different mechanism. We show that molecular binding in this case originates from quantum reflection at the steep potential drop around $\sim 1000\,a_0$.

\begin{figure}[b!]
\begin{center}
\resizebox{0.98\columnwidth}{!}{\includegraphics{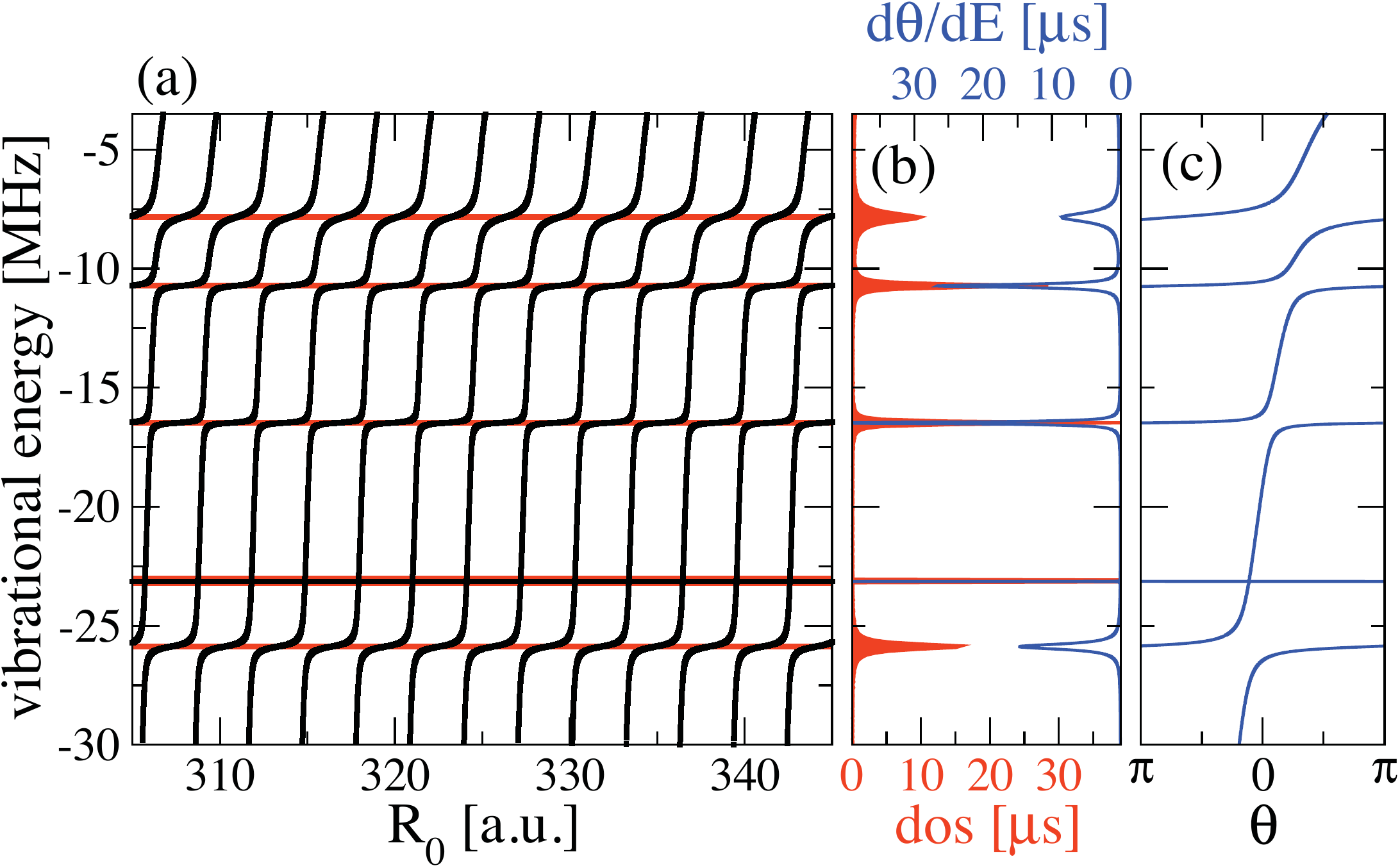}}
\caption{\label{fig3}(a) Stabilization diagram for $n=35$, showing the vibrational energies as a function of the position of the inner, reflecting boundary at $R_{0}$ (see Fig.\ref{fig2}). The corresponding density of states (dos), shown in (b), yields the energies and widths of the molecular resonances, in perfect agreement with the corresponding results obtained from the reflection phase (c).}
\end{center}
\end{figure}

Due to the enormous potential drop and flat potential energy curve at the avoided crossing, the molecular wave function behaves in the inner region very much like that of a free particle with a relatively large kinetic energy compared to the outermost part of the potential curve. The large discontinuity leads to a dramatic change in kinetic energy over a small distance and the wavefunction is reflected at this point, similar to what happens at a large impedance mismatch in an electrical circuit, and the wavefunction piles up at large $R$. Consequently, the long-range molecular structure is largely unaffected by the details of the interaction at small distances, and inaccuracies in the shorter ranged energetically deeper parts of the interaction potential can safely be neglected.
The small leakage into this region is treated as \emph{inward} scattering - in analogy to standard \emph{outward} scattering problems - such that the molecular states can be obtained from a stabilization procedure \cite{hata70}.  
Precisely, we place a reflecting boundary in the inner region of the potential well at $R_{0}$ (see inset of Fig.~\ref{fig2}) and record the energy eigenvalues as a function of $R_{0}$. The observed energies are clearly visible in the resulting stabilization plot (see Fig.\ref{fig3}a) and appear as sharp maxima in the corresponding density of states (see Fig.\ref{fig3}b).

Within an alternative approach that underlines the crucial role of quantum reflection, we consider an outgoing plane wave injected at $R=0$ with an energy below the height of the innermost well. We study the complex reflection coefficient ${\mathcal{R}}=|{\mathcal{R}}|e^{i\Theta}$ of reflection off the outer potential plateau at $R\sim 1000\,a_0$.
The scattering phase $\Theta(E)$ can be related to the time delay $\tau=\mathrm{d}\Theta/\mathrm{d}E$ between a scattered and a freely propagating particle \cite{wig55}.
In Fig.~\ref{fig3}(c) we show the phaseshift for the molecular curve of the $n=35$ Rydberg state while its derivative is depicted by the right curve in Fig.~\ref{fig3}(b).
The energy dependence of $\tau(E)$ exhibits a small background due to classical slow-down of the outgoing  particle, moving up the potential hill.
In perfect agreement with the calculated density of states (left curve in Fig.~\ref{fig3}(b)) we find an additional series of sharp resonances riding on top of this background. Consequently, these resonances are of non-classical origin and mark the occurrence of a quasi-bound state. The resonance line shape is of Breit-Wigner type, and, according to Wigner's time delay interpretation, yields the lifetime of the corresponding quasi-bound state, in this case the time the system spends in the innermost part of the potential well before penetrating into the outer part of the potential. Adopting this to the vibrational states in the outer part, $\tau$ accords to the time the ground state atom spends in the outer region before entering the inner part of the potential.

Such quasi-bound states, occur, e.g., in strong-field ionization of atoms and molecules \cite{gdb98} and affect transport phenomena in multi-layered semiconductors \cite{bue90}, where localized states are confined inside a well by a high tunneling barrier. In contrast to these systems, the long-range vibrational states studied here arise nearly entirely from quantum reflection keeping the ground state atom outside the potential drop.
To distinguish this purely quantum mechanics-based reflection from semiclassical phenomena, the spatial variation of the de Broglie wavelength $\lambda=2\pi/\sqrt{2M[E-U(r)]}$ has to be considered \cite{cht97}. A a slow variation of $\lambda$ defines the semiclassical region.
Depending on the vibrational energy, this condition is violated, i.e. $\mathrm{d}\lambda/\mathrm{d} r>1$, around the potential maxima in the outer plateau region.
Indeed the innermost of these 'badlands' extends deep into the left of the sharp potential drop.
To identify the dominant contributions of the molecular curve, we have adapted the method of \cite{gv96} to determine the reflection coefficient of the two inner potential maxima. We find, that even for the highest lying excited state for Rb(35s), the reflection coefficient is a as high as $|{\mathcal{R}}|\gtrsim0.7$.

This clearly demonstrates that these unique vibrational states arise from the special form of the interaction potential leading to quantum reflection near the deep abyss due to the avoided potential crossing that results from the p-wave scattering resonance.

\section{Conclusion}

The remarkably good agreement between experiment and theory over a range of principal quantum numbers attests to the accuracy of Fermi's original pseudopotential approach in describing interactions in highly excited multi-atom systems. A non-perturbative approach of these interactions is, however, necessary and leads to an unusual binding caused by internal quantum reflection. The latter is crucial for the existence of excited dimer states, since it prevents collapse to small interatomic distances, which would lead to fast decay of the molecule. 
Since the molecular structure is found to be sensitive to the characteristics of both s-wave and p-wave scattering, long-range Rydberg molecules are an accurate experimental platform to investigate electron-atom collision in a previously inaccessible ultralow-energy domain.

The fact that these exotic molecules have huge bond length but nevertheless a vibrational spectrum with several states make a new type of ultracold Rydberg chemistry feasible: larger polymers or even heteronuclear molecules can be realized, where the number of atoms, the constituent atomic elements and the addressed Rydberg state allow for the precise selection of the properties of the long-range molecule.
Indeed, the three-body photoionization demonstrated in this work, may be regarded as an initiating step towards the production of larger and more complex polyatomic molecules. In particular, the high sensitivity of Rydberg atoms to external fields provides an ideal starting point for coherent control and the investigation of molecular dynamics and reactions in a new regime.

\section{Methods}
\subsection{Molecular potential calculations}
Our non-perturbative potential calculations are based on the short range character of the electron-ground state atom interaction. Hence, outside the range of $V_{\rm eA}$ the electronic wavefunction 
\begin{equation}\label{expa1}
\psi({\bf r})=\sum_{lm} C_{lm}G_{lm}({\bf r},E)
\end{equation}
can be expanded in the solutions $G_{lm}=\frac{2\pi}{(2l-1)!!} [Y_{lm}(\nabla^{\prime})G_{\rm A}({\bf r},{\bf r}^{\prime},E)]_{r^{\prime}=0}$ of the atomic Hamiltonian $H_{\rm A}$. They are determined by the corresponding Green's function $G_{\rm A}$, obtained with quantum defect theory.
On the other hand, in the vicinity of the perturbing atom, we can neglect variations of the electron-core interaction [$V_{\rm eA^+}(|{\bf r}-{\bf R}|)\approx V_{\rm eA^+}(R)$]. In this case the Hamiltonian (\ref{ham}) describes a free-particle scattering from a finite-range potential \cite{hy57}, such that the wavefunction
\begin{equation}\label{expa2}
\psi({\bf r})=\sum_{lm}c_{lm}Y_{lm}({\bf n})[r^{-l-1}+...+B_l(r^l+...)]
\end{equation}
is entirely determined by the scattering phase shifts $\delta_l(k)$ via $(2l-1)!!(2l+1)!!B_l=k^{2l+1}\cot \delta_l(k)$ of the different partial waves. Finally, the electron momentum $k$ is expressed through the semi-classical relation $E=k^2/2+V_{\rm eA^+}(R)$, and its total energy $E$ is obtained by matching the two different expansions Eq.(\ref{expa1}) and Eq.(\ref{expa2}) as $r\rightarrow 0$.
\subsection{Experimental scheme}
The experiments are performed in an ultracold cloud ($T=3\mu$K) of magnetically trapped $^{87}\mbox{Rb}$ atoms in the ground state $5s_{1/2}, F=2, m_F=2$. The  $ns_{1/2}$ Rydberg state is addressed via a pulsed two-photon excitation: The first laser at 780~nm is blue-detuned with respect to the $5p_{3/2}$ state and together with the second laser at 480~nm resonant to the transition $5s_{1/2}\rightarrow ns_{1/2}$. In the experiment, the two cw-lasers with a combined linewidth  below 1~MHz are switched on for $3\mu s$. For signal detection, we apply a field ionization pulse immediately after the laser excitation and probe the ions with a microchannel-plate. Details regarding the experimental setup and the sequence for the photoassociation spectroscopy can be found in \cite{bbn09}.\\

\begin{acknowledgments}
We would like to thank I.~Fabrikant, H.~Sadeghpour and F.~Robicheaux for helpful discussions. B.~Butscher thanks the Carl Zeiss foundation and J.~P. Shaffer the Humboldt foundation for financial support. Parts of this work are funded by the Deutsche Forschungsgemeinschaft (DFG) within the SFB/TRR21 and the project PF~381/4-1.
\end{acknowledgments}
\end{document}